\documentclass[twocolumn,osajnl,graphicx]{revtex4}

\usepackage{graphicx} 
\usepackage{array}    
\usepackage{amsmath}   
\usepackage{amssymb}   
\usepackage{wrapfig}  

\begin{document}

\title{Addressing atoms in optical lattices with Bessel beams}

\author{M. Saffman}
\affiliation{ Department of Physics, University of Wisconsin, 1150 University Avenue,  Madison, Wi., 53706}

\begin{abstract}
A method of synthesizing localized optical fields with zeroes on a periodic lattice is analyzed. The applicability to addressing atoms trapped in optical lattices with low crosstalk is discussed. 
 \end{abstract}

\date{\today}

\maketitle

There is much current interest in using atoms trapped in optical lattices for quantum logic devices\cite{ref.qulattice1,ref.qulattice3}.
One of the challenges in implementing this scheme is posed by the need to address individual atoms with near resonant light. 
Current estimates of error limits for scalable quantum computing require primitive logic operations with errors as low as\cite{ref.steane, ref.preskill} $O(10^{-5}-10^{-6}).$ The corresponding limitation on optical crosstalk when addressing one atom in a lattice can only be quantified in the context of a particular choice of 
quantum gate. In atomic schemes error rates for operations that are dependent on single photon processes  tend to scale with the intensity.  This implies that  the intensity leakage of a logical control or state readout beam at a site adjacent to a site being addressed should not exceed 
$O(10^{-5}).$ In this letter I describe a novel approach to image formation that allows sites in an optical lattice to be addressed with minimal crosstalk. 

In order to emphasize the difficulty of achieving low crosstalk  in optical lattices consider the following simple scaling relationships. 
We have  a 1- or 2-D lattice lying in the $x-y$ plane that is defined by counterpropagating beams at wavelength $\lambda_f.$ Individual atoms are separated by a minimum distance of $d=\lambda_f/2$ and we wish to address them with near-resonant light of wavelength $\lambda.$ The most obvious approach to doing so involves 
focusing a beam propagating perpendicular to the plane of the lattice (along $\hat z$)  to a small spot. Assuming a Gaussian beam profile 
the intensity distribution is $I(\rho,z)=I_0\exp(-2\rho^2/w^2)$ where $\rho^2=x^2+y^2,$ 
$w^2(z)=w_0^2(1+z^2/z_R^2)$, $w_0$ is the beam waist at $z=0,$ and
$z_R=\pi w_0^2/\lambda.$ Expressing the ratio of the intensity  at a neighboring site to the on-site intensity as $\epsilon$ implies a  beam waist    $w_0=\sqrt{-1/(2\ln\epsilon)}~\lambda_f.$ For $\epsilon=10^{-5}$ this evaluates to $\tilde w_0=w_0/\lambda_f=0.21.$

\begin{figure}[!t]
\centering
\includegraphics[width=8.cm]{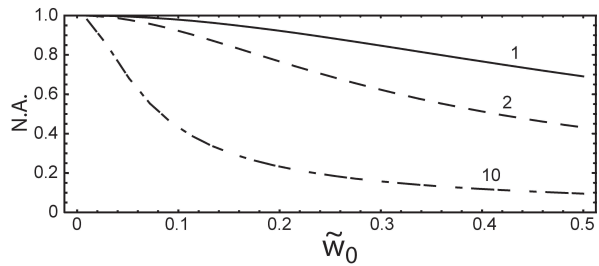}
\vspace{-.5cm}
\caption{Lens numerical aperture needed to focus a Gaussian beam to a waist $\tilde w_0$ for  $\lambda_f/\lambda=1,2,$ and $10.$  }
\label{fig.na}
\end{figure}
We can quantify the corresponding requirement on the ratio of trapping and addressing beam wavelengths in terms of the performance of high numerical aperture lenses.  We assume a lens system with aperture diameter $D$ is used to focus the addressing light. If the ratio of the lens aperture to the Gaussian waist at the aperture is given by $p=D/w(z_{\rm lens})$ then a focal plane waist of $w_0$ implies 
$D\simeq p w_0 z_{\rm lens}/z_R.$ 
A reasonable minimum requirement on $p$ to avoid rings surrounding the focal plane image is that the fraction of the optical power blocked by the aperture should not exceed the desired intensity ratio $\epsilon$ by more than a few times. Using an optimistic value of $p=3$ gives approximately 1\% transmission loss at the lens and a numerical aperture of  
\begin{equation}
NA=\frac{\left(\frac{3}{2\pi \tilde w_0}\frac{\lambda}{\lambda_f}\right)}{\left[1+\left(\frac{3}{2\pi \tilde w_0}\frac{\lambda}{\lambda_f}\right)^2\right]^{1/2}}.
\end{equation}
The variation of the lens NA with $\tilde w_0$ is shown in Fig. \ref{fig.na}. Available microscope objectives, as well as high resolution optical lithography lens systems, tend to have numerical apertures not more than $\sim 0.8$ unless oil immersion is used which is not compatible with atomic imaging inside vacuum chambers. This implies that 
the lattice light wavelength $\lambda_f$ must be several times  longer than the addressing wavelength $\lambda$ in order to address single atoms with negligible leakage to neighboring sites. 
Thus
a lattice of Rb atoms which can be manipulated with light near resonant with the D2 line at  $\lambda=0.78~\mu\rm m$ will not be individually resolvable in the context of scalable quantum logic unless $\lambda_f \gtrsim 1.5 \mu\rm m.$

Various solutions to the addressability problem are under study.
The most direct solution is to make  $\lambda_f\gg \lambda$.
Arrays of widely spaced traps that use many diffractively generated 
beams\cite{ref.ertmer} instead of lattices also fall into this category of solution. 
  Unfortunately  this is not compatible with loading from a Bose Einstein condensate 
via the Mott transition\cite{ref.mott}, although other loading schemes may still be 
used\cite{ref.sw,ref.weiss}. 
 Another possibility is to keep $\lambda\sim\lambda_f$ 
but load the lattice  so that only every few lattice sites are occupied, 
or change the angle between the lattice beams after loading so that a longer periodicity is 
obtained\cite{ref.zoller,ref.porto}.

\begin{figure}[!t]
\centering
\includegraphics[width=7.cm]{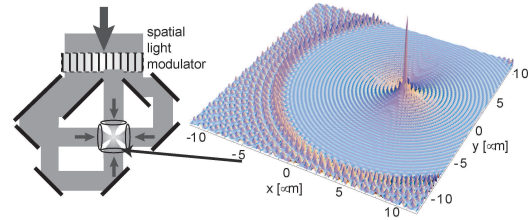}
\caption{Optical layout for writing a Bessel beam. The intensity distribution was calculated with $N=100$ and $\lambda=0.78~\mu\rm m.$ The central lobe was translated by $\Delta x=4~\mu\rm m$ and $\Delta y = 2 ~\mu\rm m$ by adding phase offsets, see the text for details. }
\label{fig.optics}
\end{figure}

Here we analyze an alternative approach to single atom addressing in optical lattices with $\lambda\sim\lambda_f.$
The idea is to accept the fact that the light cannot be localized sufficiently well, but tailor the beam profile so that the field has zeroes 
at neighboring lattice sites. 
  The basic geometry is sketched in Fig. \ref{fig.optics}. $N$ beams, each with a phase controlled by a spatial light modulator pixel,  propagate in the $xy$ plane  and are aligned to converge on the origin at $x=y=0.$ The beams necessary for creating the optical lattice can in principle be combined with the imaging beams using dichroic mirrors. 
Each converging plane wave is polarized along $\hat z$ with  amplitude $A_j\exp[i(-{\bf k}_j\cdot{\boldsymbol \rho}-\omega t+\chi_j)]+c.c.,$ where ${\bf k}_j=k(\cos\phi_j\hat x + \sin\phi_j\hat y),$ $k=2\pi/\lambda,$ $\phi_j$ is the azimuthal angle of wave $j,$ and $A_j,~\chi_j$  are adjustable  amplitudes and phases. 
As shown in the figure this can be achieved using a single one-dimensional modulator with $N$ pixels and a system of mirrors and lenses. In the limit when $N\rightarrow\infty$ and the beam amplitudes and phases are all equal the field amplitude generated on axis is just the zero order Bessel beam $J_0(k\rho).$ 

When the field is synthesized from a finite number $N$ of plane waves it is quasiperiodic in space, and rings of secondary intrerference maxima occur as can be seen in Fig. \ref{fig.optics}. The diameter of the secondary rings can be estimated from $\Delta k d_{\rm ring}\simeq 2\pi.$
Using $\Delta k= 2k/(N/2)$ gives $d_{\rm ring}\simeq N\lambda/4$, while the actual diameter  found numerically  is some 25\% larger.
Recalling the shift theorem of Fourier analysis it is straightforward to scan the spot location over a distance of up to $d_{\rm ring}$ by adding appropriate phase offsets to the incident beams, as is shown in Fig. \ref{fig.optics}. It is possible in this way to address hundreds of atoms using a multipixel modulator,  without mechanical motion of the optical system.  
  This approach to synthesis and scanning of localized optical fields also finds application in atomic lithography\cite{ref.patent}.

A Bessel beam written in this way with $\lambda=0.78~\mu\rm m$ has a central lobe with $1/e^2$ intensity radius  of $0.22~\mu\rm m.$ However the closely spaced secondary maxima of the Bessel function give crosstalk on nearby lattice sites. We describe now a method of synthesizing a beam with zeroes at regularly spaced sites of a one-dimensional lattice. An arbitrary  solution of the two-dimensional Helmholtz equation can be written as a  Fourier-Bessel  series 
\begin{equation}
A(\rho,\theta)=J_0(k\rho) + \sum_{n=1}^\infty a_nJ_n(k\rho)\exp[i n \theta]
\label{eq.fb}\end{equation} 
with the $a_n$ complex coefficients to be determined. 
We have set the zero order coefficient to unity corresponding to a normalized field with unit amplitude at the origin.
Any field of the form (\ref{eq.fb}) can be synthesized with an amplitude and phase modulator in  the geometry of Fig. \ref{fig.optics}. 
To make this explicit we note that the  Fourier transform of (\ref{eq.fb}) is 
\begin{eqnarray}
\tilde A(q,\phi)&=&\int_0^\infty d\rho~\rho 
\int_0^{2\pi} d\theta A(\rho,\theta) \exp[i\rho q \cos(\theta-\phi)]
\nonumber\\
&=& \frac{2\pi}{k}\delta(q-k)\left(1+\sum_{n=1}^\infty a_ni^n\exp[in\phi]\right).
\label{eq.transform}
\end{eqnarray}
Thus a ring of converging plane waves with wavenumber $k$ and complex amplitudes given by the term in parentheses in  (\ref{eq.transform}) provides the desired field. 
The role of the lenses shown in Fig. \ref{fig.optics} is to tilt the wave passing through each spatial light modulator pixel towards the center of the image region, but the lenses do not tightly focus each wave.    Outside of the central spot destructive interference leads to a low background light level. From the point of view of Fourier optics focusing of a beam results from interference of the constituent plane wave components, taking into account relative phase shifts due to propagation. The usefulness of the approach to image synthesis presented here is that we can directly control the amplitudes and phases of each of the plane wave components.

\begin{widetext}

\begin{table}[!t]
\centering
\begin{tabular}{c|c|c|c|c|c|c}
 $M=$& $1$ & $2$  & $3$&$4$&$5$&$6$\\
\hline
$a_2$ &$.675$ &$.715$ & $.725$& $.728$ &$.730$&$.731$ \\
$a_4$ & & $-.118$ &$-.150$ &$-.163$ &$-.170$&$-.174$\\
$a_6$ & &  &$.0406$&$.0616$ &$.0736$&$.0814$\\
$a_8$ & &  && $-.0169$&$-.0302$&$-.0401$\\
$a_{10}$ & & && &$.00778$&$.01622$ \\
$a_{12}$ & & && & &$-.003857$ \\
\hline
max$|A|^2$ &$3.0\times 10^{-3}$ &$6.1\times 10^{-4}$ &$1.9 \times 10^{-4}$& $7.1\times 10^{-5}$&$3.6\times 10^{-5}$&$3.3\times 10^{-5}$ \\
$m_{\rm max}$&$4$ & 8&11 &14 & 19&$28$\\
14 bit $ |A|^2$&$2.9\times 10^{-3}$&$6.1\times 10^{-4}$ & $2.0\times 10^{-4}$&$9.3\times 10^{-5}$ &$7.9\times 10^{-5}$&$7.7\times 10^{-5}$ \\
\hline
\end{tabular}
\caption{Bessel  coefficients and resulting maximum crosstalk which occurs at site $m_{\rm max}$  away from the origin for 
$\lambda=0.78~\mu\rm m$ and $\lambda_f=0.8~\mu\rm m.$  The last row shows the maximum error in the first 50 neighboring sites using expansion (\ref{eq.transform}) with $N=256.$}
\label{tab.error}
\end{table}

\end{widetext}

A field consisting of a finite number of terms in (\ref{eq.fb}) that is useful for  addressing a one-dimensional lattice along $\hat x$ will have $A(\rho,0)=A(\rho,\pi)$ 
which implies that $a_n=0$ for $n$ odd. Requiring that $A$ vanish at the $M$ lattice points specified by $\rho_m=m \lambda_f/2,$ 
$\theta_m=0$, for $m=1...M$ and limiting the sum in (\ref{eq.fb}) to $M$ terms gives $M$ linearly independent equations $J_0(k\rho_m)+ \sum_{n=1}^N a_{2n}J_{2n}(k\rho_m)=0.$ These equations are easily solved for the coefficients $a_{2n}.$ The coefficients decrease rapidly with Bessel order provided $\lambda_f$  is not too much less than $\lambda.$ The series coefficients, and maximum intensity crosstalk at any site not being addressed are listed in Table \ref{tab.error}.
We see that $M=6$ is sufficient to ensure a crosstalk of a few times  $10^{-5}.$ 
In an actual implementation there are several limiting factors to consider including the number of spatial light modulator pixels, and the amplitude and phase resolution of each one.  The last row  in Table \ref{tab.error} shows the crosstalk is up to several times higher than the theoretical value if we assume the field is synthesized from 256 beams equally spaced azimuthally, with 14 bit resolution in amplitude and phase modulation  for each beam. 
Additional calculations with $\lambda_f=1.0~\mu\rm m$ give crosstalk levels  roughly 10 times lower than the example in Table \ref{tab.error}.

In summary we have described a novel method of addressing atoms in periodic 1-D lattices with low crosstalk. The extension of this approach to a 2-D lattice is complicated by the fact that a 2-D lattice has angle dependent interatomic spacings. The  Fourier-Bessel expansion (\ref{eq.fb}) can still be used, but the coefficients tend to grow rapidly with $n.$ Generalizations that are suitable for a 2-D geometry are currently under investigation.

This work was  supported by  the U. S.
Army Research Office under contract DAAD19-02-1-0083,
 NSF, and the A. P. Sloan Foundation.

\end{document}